\documentclass[3p,twocolumn]{elsarticle}
\pdfoutput=1

\journal{Journal of \LaTeX\ Templates}

\begin{document}

\begin{frontmatter}

\title{A compact imaging system with a CdTe double-sided strip detector for non-destructive analysis using negative muonic X-rays}

\author[tokyounivers,isas]{Miho Katsuragawa}
\ead{miho-k@astro.isas.jaxa.jp}
\author[kek]{Motonobu Tampo}
\author[kek]{Koji Hamada}
\author[isas]{Atsushi Harayama}
\author[kek]{Yasuhiro Miyake}
\author[tokyounivers,isas]{Sayuri~Oshita}
\author[isas]{Goro Sato}
\author[isas,tokyounivers]{Tadayuki Takahashi}
\author[oist]{Shin'ichiro Takeda}
\author[isas,tokyounivers]{Shin Watanabe}
\author[tokyounivers,isas]{Goro~Yabu}

\address[tokyounivers]{Department of Physics, University of Tokyo, 7-3-1 Hongo, Bunkyo, Tokyo 113-0033, Japan}
\address[isas]{Institute of Space and Astronautical Science, Japan Aerospace Exploration Agency, 3-1-1 Yoshinodai, Chuou, Sagamihara, Kanagawa 252-5210, Japan}
\address[kek]{High Energy Accelerator Research Organization, 2-4 Shirane Shirakata, Tokai-mura, Naka-gun, Ibaraki 319-1195, Japan}
\address[oist]{Okinawa Institute of Science and Technology Graduate University, 1919-1 Tancha, Onna-son, Kunigami-gun, Okinawa 904-0495, Japan}

\begin{abstract} 
A CdTe double-sided strip detector (CdTe-DSD) is an ideal device for imaging and spectroscopic measurements in the hard X-ray range above 10 keV. 
Recent development enables us to realize an imager with a detection area of $\sim$10 cm$^2$. An energy resolution of 1--2 keV (FWHM) and a position resolution of a few hundred $\mu$m are available from the detector.
This type of imager has been long awaited for non-destructive elemental analysis, especially by using negative muons, because energies of characteristic X-rays from muonic atoms are about 200 time higher than those from normal atoms.
With the method that uses negative muons, hard X-ray information gives the spatial distribution of elements in samples at a certain depth defined by the initial momentum of the muon beam. 
In order to study three-dimensional imaging capability of the method, we have developed a compact imaging system based on CdTe-DSD and a $\rm \phi$ 3 mm pinhole collimator as the first prototype. 
We conducted experiments with samples which consist of layers of Al, BN and LiF irradiated by negative muon beams in MUSE/J-PARC and successfully reconstruct hard X-ray images of muonic X-rays from B, N and F at various depths.
\end{abstract}


\end{frontmatter}


\section{Introduction}
Non-destructive elemental analysis using negative muons is a new and attractive method to study the composition of material samples without any damage \cite{daniel1984, kubo2008}.
When a negative muon is injected in a material, it is captured by an atom and forms a muonic atom. 
Characteristic X-rays are emitted as muonic X-rays in the cascade transition between orbits.
Since the mass of a muon (105.7 MeV/$\rm c^2$) is 207 times heavier than that of an electron, its orbit is much closer to the atomic nucleus than electrons. 
Therefore, the energy of the characteristic X-rays is also about 200 times higher. 
Even for light elements such as Li and Be, the corresponding muonic X-rays range in the hard X-ray wavelengths, above 10 keV.
Emissions of such high energy photons enable us to explore the deep interior of samples with small absorption even for light elements. 
However, it is only recently that the high intensity beam line, which is required for precise elemental analysis, has become available \cite{miyake2009}.

MUon Science Establishment (MUSE) is the world's strongest pulsed muon beam facility in Japan Proton Accelerator Research Complex (J-PARC). 
Negative muons are generated by irradiating a graphite target with a 3 GeV, 1 MW proton beam.
Since it is difficult for negative muons to be collimated, the use of a scanning method, like an electron beam, is difficult for spatially resolved elemental analyses. 
Recently, an analysis of one-dimensional depth profile was demonstrated with Ge semiconductor detectors by changing momenta of muons at MUSE \cite{terada2014, ninomiya2015}. 
However, in order to study the real distribution of elements in samples, two-dimensional spatial information, in conjunction with spectra, is indispensable. 
If we combine the image with the depth profile, we will be able to achieve a new method of three-dimensional (3-D) imaging by using negative muons.

Cadmium telluride (CdTe) has large atomic numbers (48, 52) and high density (5.8 g/$\rm cm^3$) and is suitable for hard X-ray detections. 
Based on high-resolution Schottky CdTe device \cite{takahashi1999}, we have established technologies of making a large area CdTe double-sided strip detectors (CdTe-DSDs) \cite{watababe2009, sato2016}, after a decade of research and development.
In this paper, we report the performance of a compact hard X-ray imager system with a CdTe-DSD as the first prototype.
Results of non-destructive elemental analysis using negative muons performed at MUSE/JPARC in 2014 and 2016 are also presented.


\section{Prototype of imaging system}

\begin{figure}[t]
  \begin{center}
    \includegraphics[width=70mm]{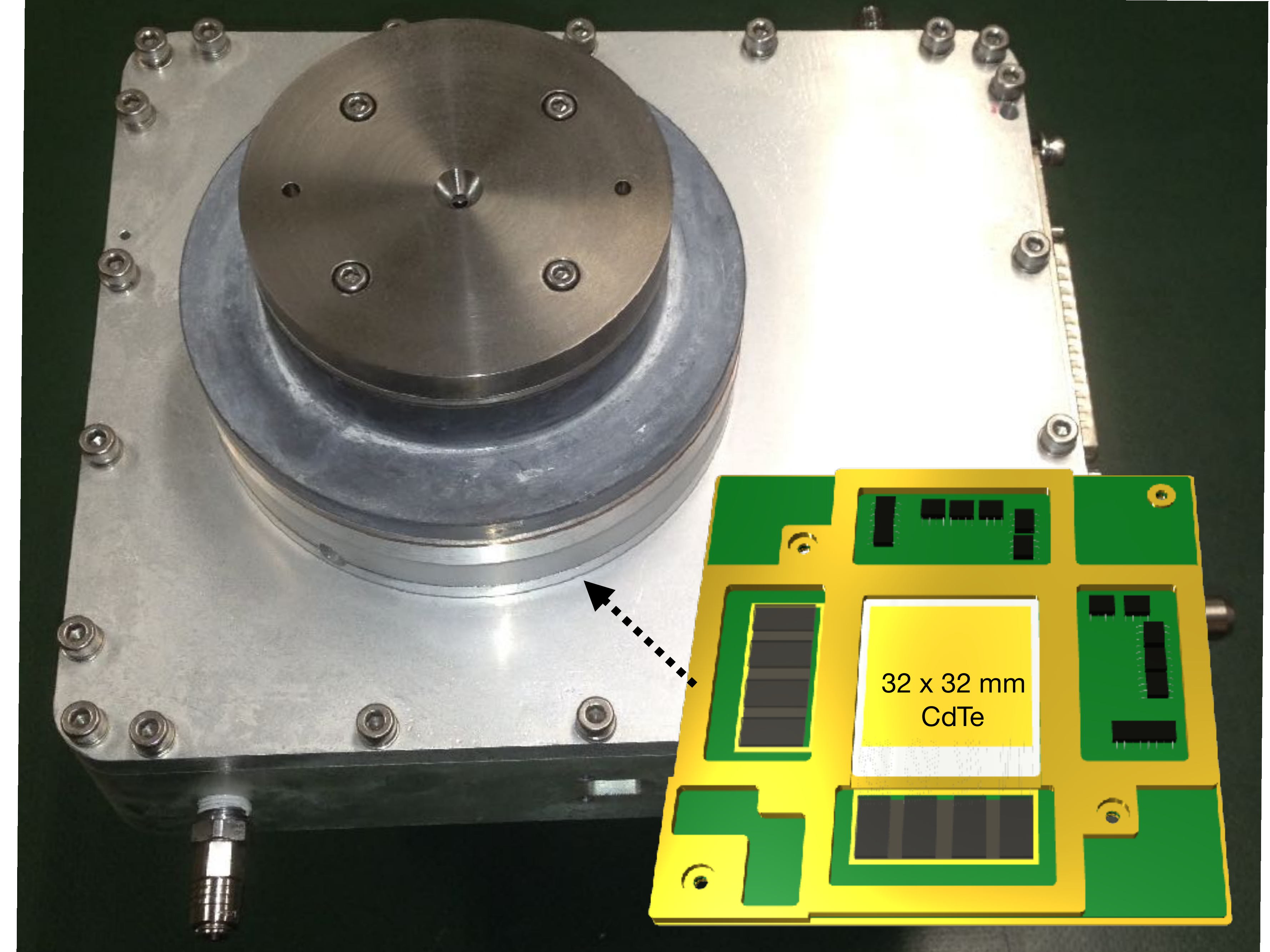}
  \end{center}
  \caption{A prototype of an imaging system with a 250 $\mu$m pitch CdTe-DSD (bottom right) and a $\phi$3 mm pinhole collimator. }
  \label{fig:imager}
\end{figure}

Figure \ref{fig:imager} shows a photo of the imager system developed for the experiment.
The system consists of one layer of CdTe-DSD, readout electronics, a pinhole collimator and an Al housing.
In order to suppress background gamma-rays, which come from outside of the pinhole FOV, Pb and Sn-Cu graded-Z shield are installed in the housing.
The CdTe-DSD has dimensions of 3.2 $\times$ 3.2 cm$^2$ and a thickness of 750 $\mu$m, providing a detection efficiency of 90\% at 60 keV.
Signals from each strip are amplified and converted to digital pulse height in dedicated analog Application Specific Integrated Circuits (ASICs) \cite{watanabe2014}.
The CdTe-DSD covers an energy range of 5--200 keV with an energy resolution of ~1.6 keV (FWHM) at 60 keV.
The detector is operated at $\rm -20^\circ C$ cooled by a Peltier cooler with a cooling power of 55 W.

In order to obtain accurate information about the distribution of elements in samples, high contrast of images is of importance. 
We adopt pinhole optics, because it provides direct imaging and is therefore almost free from image reconstruction artifacts. 
The Tungsten-made pinhole collimator we adopted for the prototype has ``knife-edge" geometry providing a field of view (FOV) of 50$^\circ$, and a 3 mm diameter hole. 
By using a plate of tungsten with a thickness of 8 mm, we ensure that the transmission of gamma-ray outside is as low $\rm 10^{-3}$.

\section{Experiment}

\begin{figure*}[t]
  \begin{center}
    \includegraphics[width=130mm]{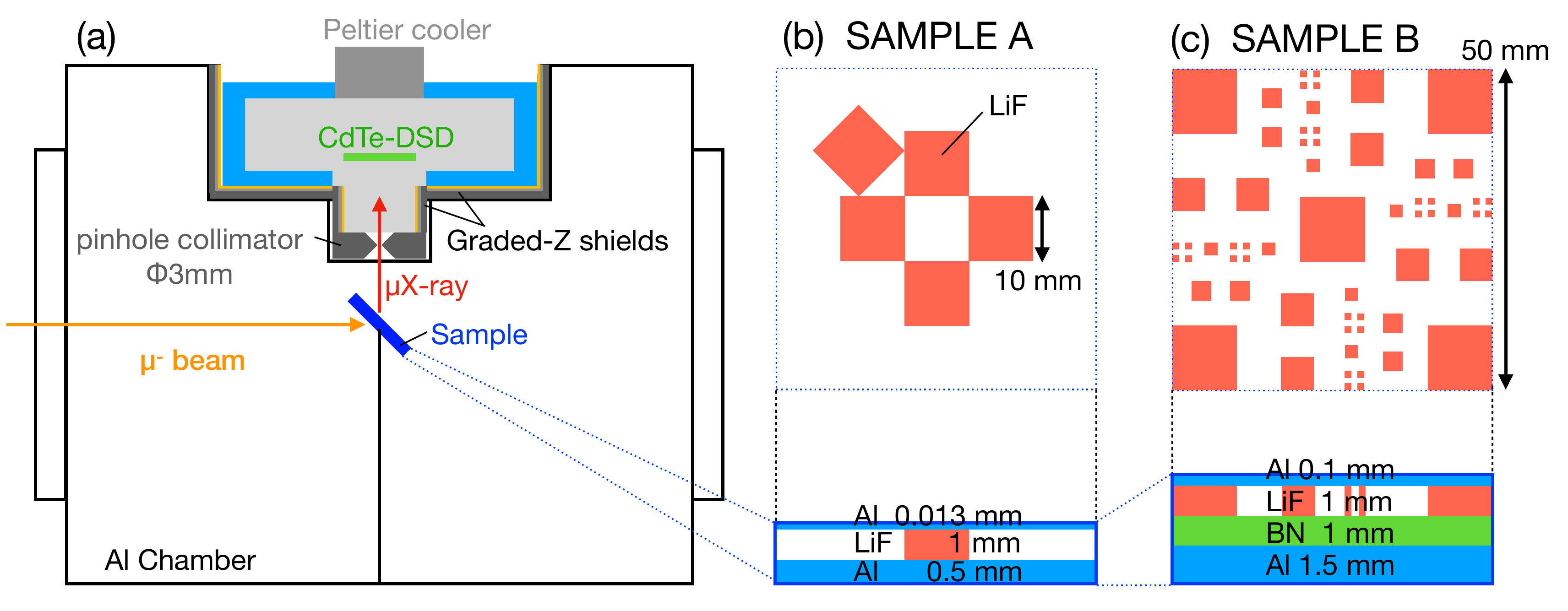}
  \end{center}
  \caption{(a) An experimental setup at D2 area/MUSE. SAMPLE A (b) and B (c) are located at the center of an  Al chamber with an inclination of 45$^\circ$.}
  \label{fig:setup}
\end{figure*}

We carried out the experiments at D2 area in MUSE/J-PARC in Dec. 2014 and April 2016.
As shown in Fig. \ref{fig:setup}(a), the imager system is installed on the top of an aluminum (Al) vacuum chamber.
A sample is set at the center of the chamber, oriented at 45 degrees to the beam. 
The distance from the sample to the pinhole is 100 mm and to the CdTe-DSD detector is 171 mm, respectively.

In order to verify the performance of the imaging system, two samples (SAMPLE A and B), which contain light elements, were prepared.
SAMPLE A has a simple structure and consists of five blocks of lithium fluoride (LiF) crystal (1.0 $\rm mm^t$) sandwiched by Al plate (0.5 $\rm mm^t$) and foil (0.013 $\rm mm^t$) (Fig. \ref{fig:setup}(b)).
On the other hand, SAMPLE B has more complex structure for the study of 3-D imaging.
It has a two-layered structure sandwiched by Al plates with 1.5 and 0.1 mm thick.
The second layer is made of boron nitride (BN) with dimensions of BN is 50 $\times$ 50 $\times$ 1.0 $\rm mm^3$.
The third layer is prepared to study how small targets can be identified.
For this purpose, LiF crystals with dimensions of 1, 2, 3, 5 and 10 mm square and 1.0 $\rm mm^t$ are arranged on the sample (Fig. \ref{fig:setup}(c)).

Adjustment of the beam momentum is very important to stop negative muons at the desired depth.
In order to find an optimum range of momenta which covers the thickness of the target, we simulate interactions of negative muons in the targets by using TRIM (TRansport of Ions in Matter) \cite{ziegler1985}.
In the simulation, the momentum distribution of incident negative muons is assumed to be a gaussian distribution with a sigma of 5\% of the mean.
A set of momenta of 40 MeV/c, 45 MeV/c and 50 MeV/c are adopted based on the expected number of stopping muons in the samples.
In this energy range, the facility provide  $\rm 4\times10^4$ negative muons per pulse in the 25 Hz beam operation.
FWHM of beam widths are about 60 mm and 70 mm in vertical and horizontal directions, respectively.

\section{Results and discussion}

\begin{figure}[ht]
  \begin{center}
    \includegraphics[width=80mm]{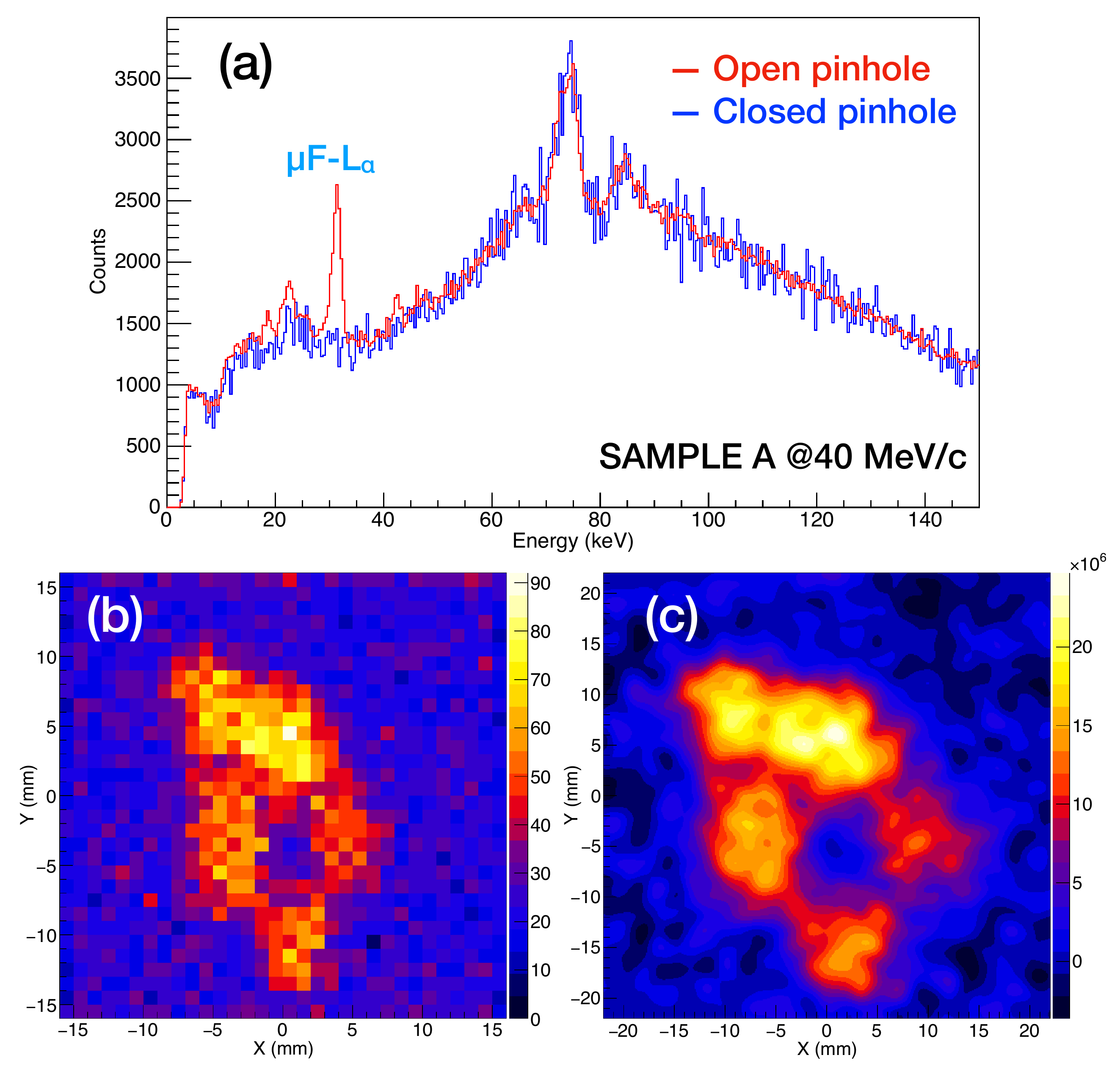}
  \end{center}
  \caption{(a) Energy spectra of SAMPLE A for the case of pinhole-open (red) and pinhole-closed (blue) at 50 MeV/c. (b)An image of $\mu$F-L$_\alpha$ (31.5 keV) and (c) an image projected onto a sample position by taking an inclination of 45$^\circ$ into account.}
  \label{fig:target_a}
\end{figure}

With the muon exposure with 20 hours at 40 MeV/c, we obtained energy spectra and an image of muonic X-rays from F ($\mu$F-L$_\alpha$) from SAMPLE A.
Fig. \ref{fig:target_a}(a) shows energy spectra for the case of pinhole-open and pinhole-closed. 
When we open the pinhole, a line of $\mu$F-L$_\alpha$ (31.5 keV) is clearly detected, the line disappears when we close the pinhole. 
While, the continuum component is same for both cases.
These results indicate that muonic X-rays of $\mu$F-L$_\alpha$ actually come from the target and the continuum component is likely due to bremsstrahlung emission from electrons, leaked through the collimator and the housing.
The 74.9 keV and 84.9 keV lines are fluorescent X-rays from Pb (K$_\alpha$ and K$_\beta$), because a part of the shield is made of Pb only.
An image of $\mu$F-L$_\alpha$ (Fig. \ref{fig:target_a}(b)) clearly demonstrates that the imaging system we have developed reveals the distribution of material hidden by Al layers. 
Fig. \ref{fig:target_a}(c) shows an image projected onto a sample position by taking an inclination of 45$^\circ$ and the effects of distance into account.
The image is smoothed by a gaussian with FWHM of 2 mm.
The arrangement of all LiF crystals with dimension of 10 mm square is clearly seen and agrees with the actual distribution of LiF on SAMPLE  A (Fig. \ref{fig:setup}(b)).

\begin{figure}[t]
  \begin{center}
    \includegraphics[width=80mm]{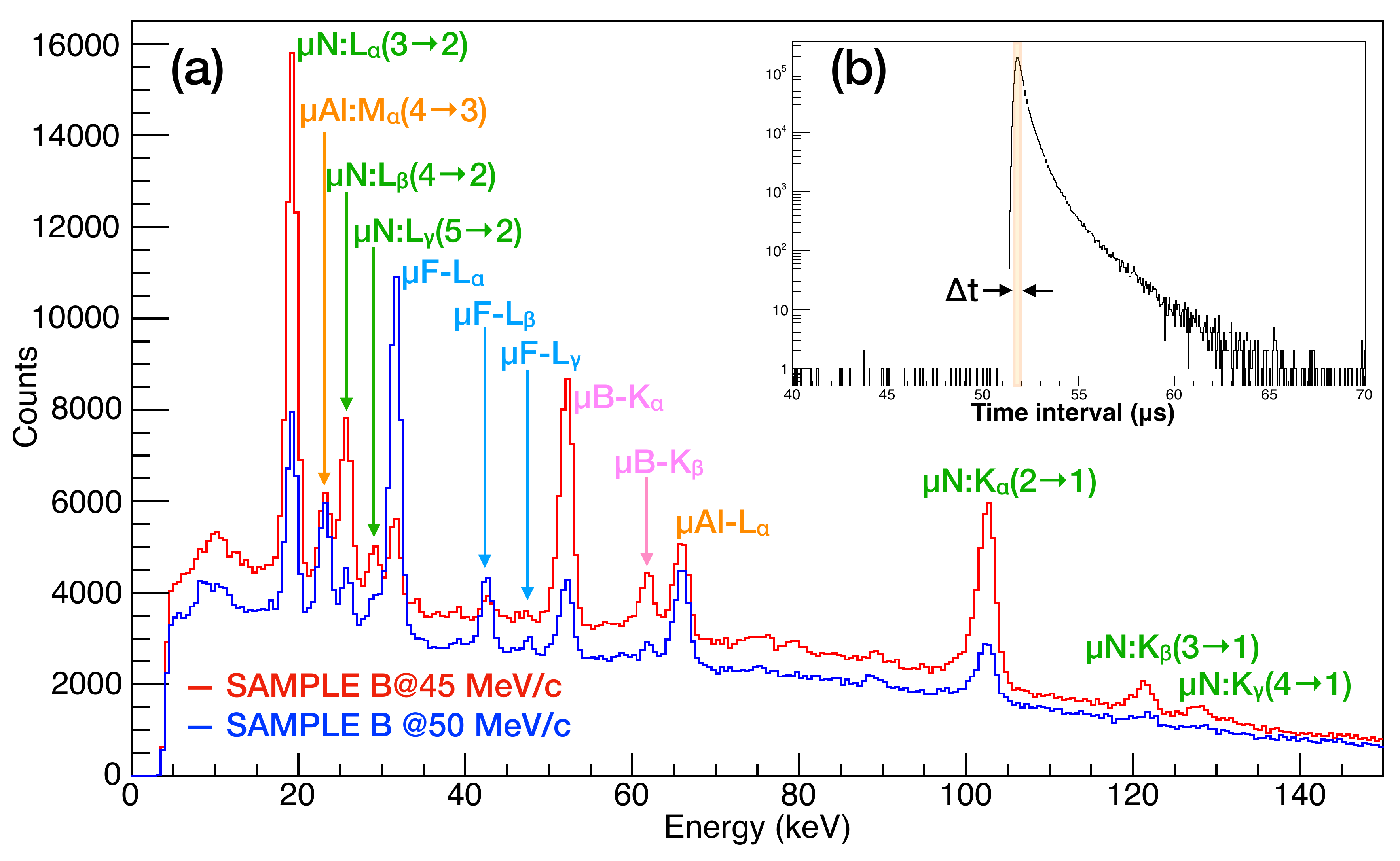}
  \end{center}
  \caption{(a) Energy spectra of SAMPLE B at 45 MeV/c (red) and 50 MeV/c (blue). (b) The time distribution of the interval between the signal synchronized with each beam pulse and the trigger pulse from detector. }
  \label{fig:target_b}
\end{figure}

As the next step, we improved the imaging system to reduce continuum background seen for SAMPLE A. 
Graded shields, consisting of Sn and Cu plates, are installed inside the Pb shield to block fluorescent X-rays from Pb.
Also, we introduced a new signal for offline analysis, which is generated from the beam line electronics and synchronized with the start timing of each beam pulse within 2 ns.
The time interval between the signal and the trigger pulse from the detector is recorded.
Fig. \ref{fig:target_b}(b) shows the time distribution of the interval.
Events distributed in the tail component are considered to be background.
By changing the range of the interval ($\Delta$t), the continuous component in the spectrum is reduced.
When we use $\rm \Delta t =$ 400 ns, the signal to background ratio becomes maximum without losing the line flux.

In order to try an advanced 3-D visualization, we use SAMPLE B, which has more complex structure than SAMPLE A.
Figure  \ref{fig:target_b}(a) shows energy spectra obtained at a beam momentum of 45 MeV/c and 50 MeV/c. 
The exposure time is 22 hours for each momentum setup.
As shown in the figure, we succeed to identified many lines of muonic X-rays.
The intensity of lines changes when we change the momentum of muons. 
These changes are consistent with the results obtained from the simulation by TRIM.
Because of the smaller atomic number of Li than F, the intensity of muonic X-rays of Li is smaller than that of F.
In addition, the line of $\mu$Li-K$_\alpha$ (18.7 keV) appears too closely to distinguish it from the line of $\mu$N-L$_\alpha$ at 19.0 keV.

\begin{figure}[t]
  \begin{center}
    \includegraphics[width=80mm]{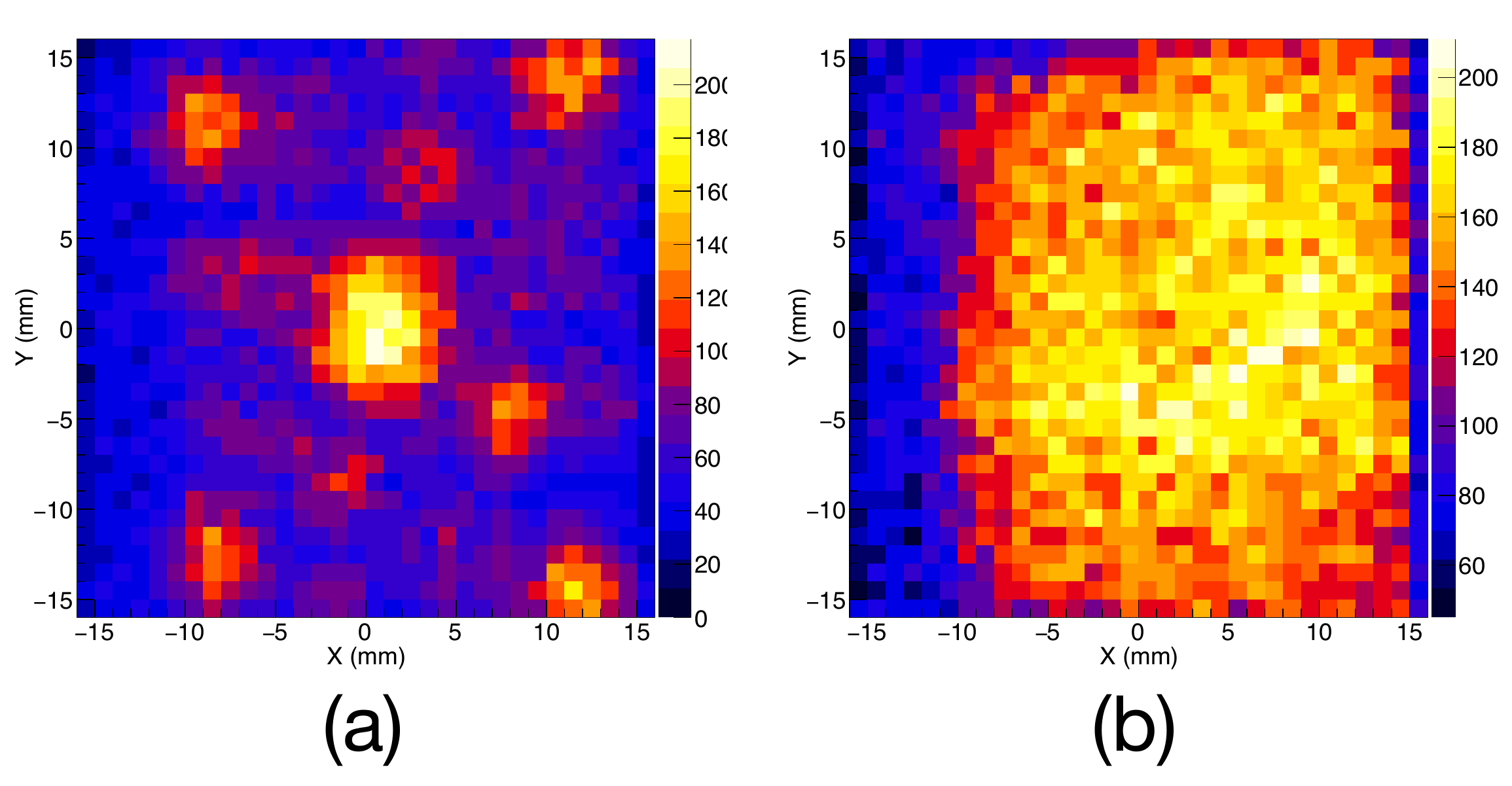}
  \end{center}
  \caption{(a) An image of F using a line of $\mu$F-L$_\alpha$ (31.5 keV) at 50 MeV/c. (b) An image of BN plate using lines of $\mu$N-L$_\alpha$ (19.0 keV) and $\mu$B-K$_\alpha$ (52.3 keV) at 45 MeV/c.}
  \label{fig:image1}
\end{figure}

Fig. \ref{fig:image1}(a) shows an image of SAMPLE B using the line of $\mu$F-L$_\alpha$ at 50 MeV/c.
Some 5 mm square crystals are also seen in the image, in addition to the five crystals with 10 mm square.
The image of the BN plate using lines of $\mu$N-L$_\alpha$ and $\mu$B-K$_\alpha$ (52.3 keV) at 45 MeV/c shown in Fig. \ref{fig:image1}(b) is a flat image as expected from the structure of the sample.

\section{Conclusion}
In order to demonstrate a new method in non-destructive analysis, which is enabled by a great combination of hard X-ray imaging technology and high-intensity negative muon beam, we conducted two steps of experiments with two samples in MUSE/J-PARC D2 beam line.
We prepared the SAMPLE A for a proof of concept of the analysis, and SAMPLE B for advanced 3-D visualization.

The role of the first experiment is to clearly show the principle of the analysis, in a word, the visualization of distribution of light elements hidden behind an optically thick surface. 
Indeed, the LiF sample is shielded by Al with a thickness of 13 $\mu$m, which is a fatally thick obstacle for characteristic X-rays (0.676 keV) from normal Fluorine. 
The combination of the characteristic ``hard" X-rays from muonic Fluorine (31.5 keV) and high-resolution CdTe imaging detectors, for the first time, enables us to see the interior distribution of light atoms through on an Al shield. 
We clearly justified our imaging result by means of the open/close measurement of pinhole optics. 

The result provided by the first experiment suggests the possibility of 3-D mapping of light elements inside a sample since the depth where an incident muon stops in the sample can be controlled by choosing the momentum of beam.
We designed SAMPLE B consisting of two layers of material structures, BN and LiF, for the feasibility study of the 3-D mapping by means of this depth-scanning-with-imaging method. 
Even though the scanning points are limited to two points, 45 MeV/c and 50 MeV/c, it is clear that the structure of elements (F, B and N) in each layers in the sample is detected.
This results demonstrate that our approach is working well. 
Quantifying the position determination accuracy along with depth direction needs data given by finer momentum scanning, and that is the next step of our development.

\section*{Acknowledgment}
This work was supported in part by Photon and Quantum Basic Research Coordinated Development Program as the Ministry of Education, Culture, Sports, Science and Technology (MEXT) scientific grants, Japan Society for the Promotion of Science (JSPS) KAKENHI grant number 16H02170, 16H03966, 24105007 and 26706024. MK is financially supported by a Grant-in-Aid for JSPS Fellows.

\section*{References}

\bibliography{mybibfile.bib}

\end{document}